\documentclass[sigconf, screen]{acmart}

\usepackage{balance}
\usepackage{graphicx, caption}
\usepackage[most]{tcolorbox}

\AtBeginDocument{%
  \providecommand\BibTeX{{%
    \normalfont B\kern-0.5em{\scshape i\kern-0.25em b}\kern-0.8em\TeX}}}

\setcopyright{acmlicensed}
\acmPrice{15.00}
\acmDOI{10.1145/3617555.3617875}
\acmYear{2023}
\copyrightyear{2023}
\acmSubmissionID{fsews23promisemain-p12-p}
\acmISBN{979-8-4007-0375-1/23/12}
\acmConference[PROMISE '23]{Proceedings of the 19th International Conference on Predictive Models and Data Analytics in Software Engineering}{December 8, 2023}{San Francisco, CA, USA}
\acmBooktitle{Proceedings of the 19th International Conference on Predictive Models and Data Analytics in Software Engineering (PROMISE '23), December 8, 2023, San Francisco, CA, USA}
\received{2023-07-07}
\received[accepted]{2023-07-28}
\begin{document}

\title{BuggIn: Automatic Intrinsic Bugs Classification Model using NLP and ML}



\author{Pragya Bhandari}
\affiliation{%
  \institution{University of British Columbia, Okanagan}
  \city{Kelowna}
  \state{British Columbia}
  \country{Canada}}
\email{pragya18@mail.ubc.ca}

\author{Gema Rodríguez-Pérez}
\affiliation{%
  \institution{University of British Columbia, Okanagan}
  \city{Kelowna}
  \state{British Columbia}
  \country{Canada}}
\email{gema.rodriguezperez@ubc.ca}


\begin{abstract}
Recent studies have shown that bugs can be categorized into intrinsic and extrinsic types.  Intrinsic bugs can be backtracked to specific changes in the version control system (VCS), while extrinsic bugs originate from external changes to the VCS and lack a direct bug-inducing change. Using only intrinsic bugs to train bug prediction models has been reported as beneficial to improve the performance of such models. However, there is currently no automated approach to identify intrinsic bugs. To bridge this gap, our study employs Natural Language Processing (NLP) techniques to automatically identify intrinsic bugs. Specifically, we utilize two embedding techniques, seBERT and TF-IDF, applied to the title and description text of bug reports. The resulting embeddings are fed into well-established machine learning algorithms such as Support Vector Machine, Logistic Regression, Decision Tree, Random Forest, and K-Nearest Neighbors. The primary objective of this paper is to assess the performance of various NLP and machine learning techniques in identifying intrinsic bugs using the textual information extracted from bug reports. The results demonstrate that both seBERT and TF-IDF can be effectively utilized for intrinsic bug identification. The highest performance scores were achieved by combining TF-IDF with the Decision Tree algorithm and utilizing the bug titles (yielding an F1 score of 78\%). This was closely followed by seBERT, Support Vector Machine, and bug titles (with an F1 score of 77\%). In summary, this paper introduces an innovative approach that automates the identification of intrinsic bugs using textual information derived from bug reports.

\end{abstract}





\maketitle

\section{Introduction}
Software bugs have been the nemesis of software developers, persisting in a recurring cycle wherein a bug is introduced during the development process through code modifications. 
It has conventionally been assumed that all bugs originate from a developer's faulty lines of code. Nevertheless, recent studies have revealed that this assumption does not always hold true~\cite{rodriguez2020bugs,Gema_origin2018}. With the rising usage of external libraries, dependencies, and APIs in a project, 
there have been scenarios wherein the bugs are caused by a change in the environment where the software is used, or because requirements changed in an external library used by the project, or by an external change to the version control system (VCS) of the project~\cite{Gema_origin2018, rodriguez2020bugs}.

\begin{figure}[ht]
    \centering
    \includegraphics[width=1\linewidth]{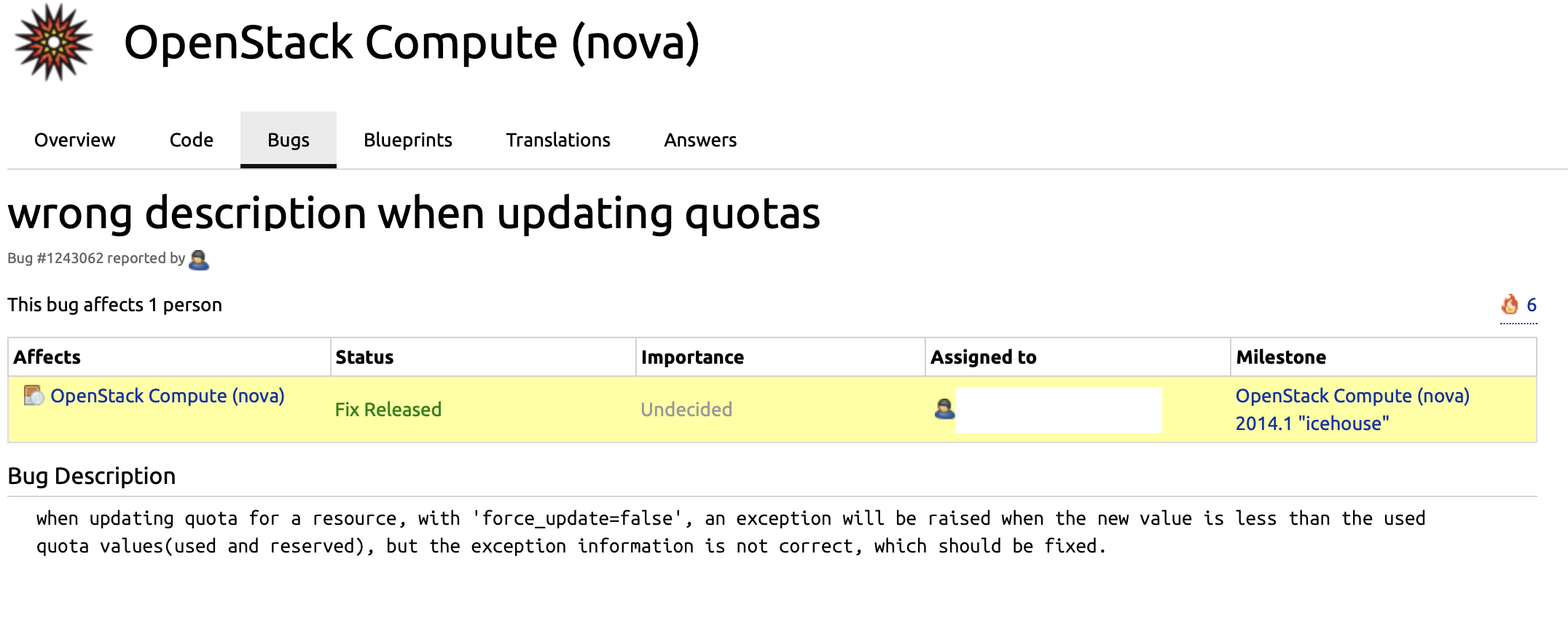}
     \caption{Bug report of an intrinsic bug} 
     \label{fig:a}
\end{figure}

Consequently, bugs can be classified into two types based on the origin of their causation: intrinsic and extrinsic. Intrinsic bugs can be traced back to a specific point within the VCS where the bug-causing change was made. Figure~\ref{fig:a} reports that the exception information provided when a specific scenario happens is inaccurate and, therefore, requires correction. According to the definition of intrinsic bugs~\cite{Gema_origin2018}, the bug report in Figure~\ref{fig:a} is clearly describing an intrinsic bug, as the bug originated in a change in the source code of the project. That source code was buggy and can be identified in the project's VCS.

\begin{figure}[ht]
    \centering
    \includegraphics[width=1\linewidth]{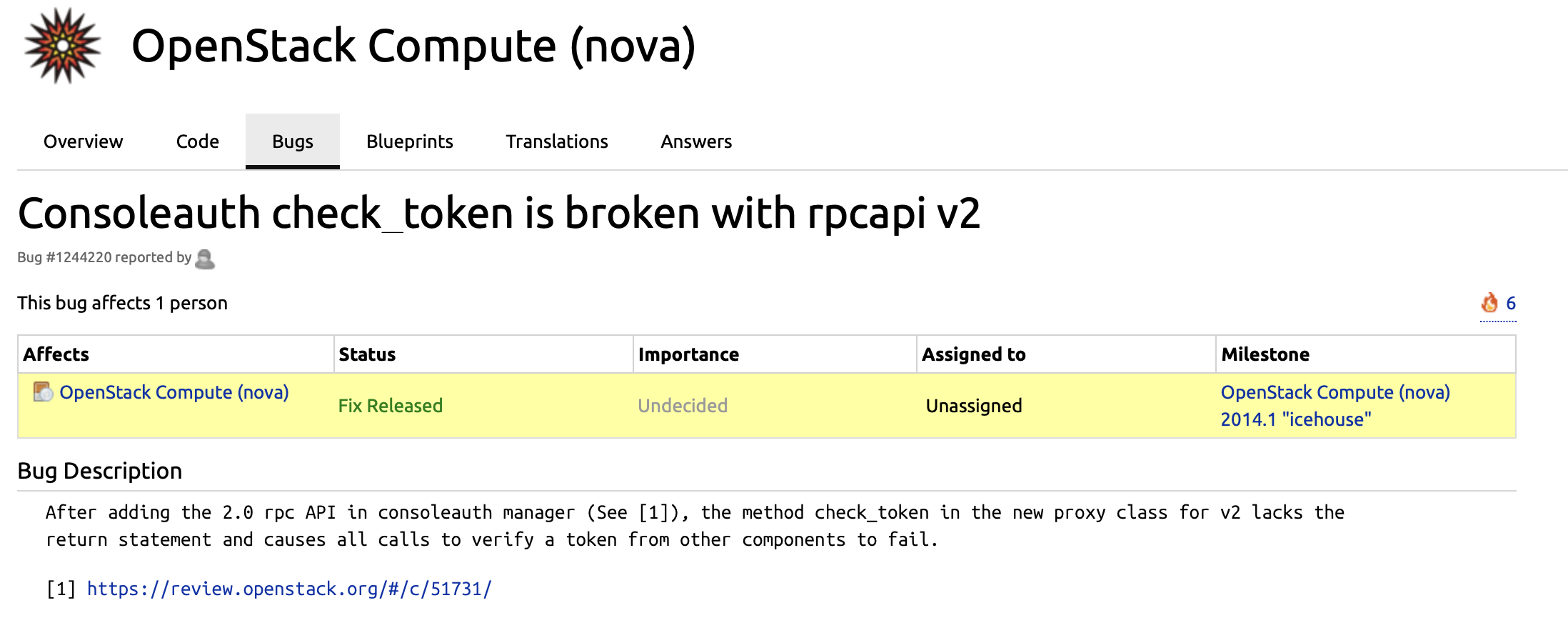}
    \caption{Bug report of an extrinsic bug} 
    \label{fig:b}
\end{figure}

Extrinsic bugs are instigated by changes that lie beyond the project's domain and control, making it impossible to trace them back to a specific bug-inducing commit in the VCS. Figure~\ref{fig:b} describes a bug in a method after incorporating the version 2.0 rpc API into the system. Consequently, this omission results in the failure of all calls made by other components. According to the definition of extrinsic bugs~\cite{Gema_origin2018}, the bug report in Figure~\ref{fig:b} is clearly describing an extrinsic bug, as the bug was not caused by buggy source code but by a change in a third-party API. That change cannot be found in Nova's VCS therefore, it's not intrinsic to Nova.

Extrinsic and intrinsic bugs pose a significant challenge to the outcomes of bug prediction models that rely on learning from past bug-inducing commits to forecast future instances as such models have consistently treated all bugs as intrinsic. For example, Just-In-Time (JIT) bug prediction models~\cite{JIT} enable the identification of potentially risky commits during the integration of code changes into the VCS. JIT models allow practitioners to allocate their testing resources effectively, focusing on reviewing and addressing the most critical commits while the changes are still fresh in the minds of the authors~\cite{JIT}. 

The accuracy of bug prediction models heavily depends on identifying the specific commit that introduced the bug. Previous studies have utilized algorithms such as SZZ~\cite{sliwerski2005changes}, V-SZZ~\cite{bao2022v}, or RA-SZZ~\cite{neto2018impact} to create bug corpus datasets and link those bugs with their bug-inducing commits. Unfortunately, for extrinsic bugs, there is no way to identify the associated bug-inducing commit in the VCS~\cite{Gema_origin2018}. Consequently, when employing these approaches, the dataset fed into bug prediction models becomes susceptible to noise, as extrinsic bugs may be mistakenly linked to bug-inducing commits. Recent work demonstrated that excluding extrinsic bugs from the dataset and training the bug prediction model exclusively on intrinsic bugs significantly enhances its performance~\cite{Gema_watchout}. 

However, currently, the only approach to classify bugs as intrinsic or non-intrinsic requires manually analyzing the bugs and their textual information~\cite{Gema_origin2018, rodriguez2020bugs}, which is a very labor-intensive task. Therefore, no present studies have tried to train JIT models with only intrinsic bugs. To bridge this gap and assist researchers and practitioners in curating reliable datasets to improve the performance of bug prediction models, we present the \emph{first automated approach for the identification of intrinsic bugs}. As suggested in previous studies~\cite{B15,B28}, we leverage the rich information contained within the natural language of bug reports, coupled with advanced Natural Language Processing (NLP) techniques and Machine Learning (ML) algorithms, to enable the effective identification of intrinsic bugs.

Hence, we used a manually labeled dataset of bug reports that were categorized as intrinsic, extrinsic, and non-bugs~\cite{Gema_watchout}. Then we designed a series of experiments that involve three-fold comparisons using the textual features from the title and description of bug reports, two embedding techniques (seBERT \cite{seBERT} and TF-IDF \cite{TF-IDF}), and five ML algorithms. Our goal is to assist researchers and practitioners in collecting reliable dataset that only contains intrinsic bugs, i.e., bugs for which a bug-inducing commit can be identified. To achieve our goal, this study evaluates the efficacy of different embedding techniques and ML models in identifying intrinsic bugs based on the textual information extracted from bug reports.

Our findings reveal interesting distinctions in the performance of seBERT and TF-IDF based on the characteristics of the text being analyzed. While seBERT obtained higher performance when processing larger and more verbose bug report descriptions, TF-IDF yields better results when handling concise bug report titles. Taking into account the broader perspective of the automatic bug classification model's best results, we observe that utilizing bug report titles as independent variables, employing TF-IDF for embedding, and employing the Decision Tree classifier model lead to an exceptional overall F1-score of 78\%. However, it is worth noting that when examining the differences in F1-scores between seBERT and TF-IDF, we observe a maximum discrepancy of merely one point in both cases where either seBERT outperforms TF-IDF or vice versa. This indicates that both embedding techniques showcase comparable performance for our specific dataset. 

The main contributions made by this paper are:
\begin{itemize}
    \item The first classification that automatically categorizes reported bugs as intrinsic, solely relying on the textual information provided.
    \item A comprehensive experimental evaluation to assess the performance of various NLP and ML techniques in classifying bug reports as intrinsic or non-intrinsic.
    \item A replication package\footnote{https://zenodo.org/record/8125762} that can be used by the research community to replicate our study and validate our results.
\end{itemize}

The remainder of this paper is organized as follows. Section~\ref{related} discusses related work. Section~\ref{approach} introduces  the details of our approach. Section~\ref{experiments} describes the experimental setup used. Section~\ref{results} presents the results. Section~\ref{discussion} discusses the findings and implications, while Section~\ref{threats} contains the threats to their validity. Finally, Section~\ref{conclusion} draws conclusions and provides the link to our reproducibility package.

\section{Related Work}
\label{related}
Numerous research studies have delved into the potential of leveraging bug report text for bug classification tasks, employing diverse approaches tailored to different classification criteria. While a significant portion of the existing bug classification literature has predominantly focused on determining whether a given bug report represents an actual bug or not \cite{B10, B12, B17, B18, B21, B22}, other studies have built models to predict bug report priority or severity level~\cite{B4, B13, B28, B9}, to identify the root cause of a bug~\cite{B15, B20, B14, B28}, and to predict GitHub issue labels such as Bug, Enhancement, or Question~\cite{B2, B5, B13}. 

This section discusses the existing literature on using textual information from bug reports to classify software issues.


\subsection{Bug Categorization}
Li et al.~\cite{B10} utilized bug title and description as input features, employing Word2Vec for embedding generation and an Attention-based Bi-directional Long Short Term Memory (ABLSTM) as the classification model to categorize bug reports as either bugs or non-bugs. Moreover, Qin and Sun \cite{B12} tackled the same classification criterion using bug summary and description. They compared the performance of an Long-Short Term Memory (LSTM) model with a softmax layer, both with and without pre-trained word embeddings from Google \cite{Google_pretrained}, against a Naive Bayes (NB) classifier combined with a topic-based Latent Dirichlet Approach (LDA)~\cite{LDA} and a Random Forest (RF) classifier combined with an n-gram Inverse Document Frequency (IDF) based approach \cite{B22}. Their results indicate that the LSTM model with a final softmax layer, without the use of pre-trained word embeddings, achieved the most favorable results~\cite{B12}.

Zeng et al~\cite{B17} aimed to classify bugs into bugs and non-bugs by comparing a cross-project bug classification learning technique with a project-wise classification baseline. Their approach involved utilizing the concatenated form of bug titles and descriptions. Support Vector Machine (SVM) and Logistic Regression (LR) were employed within their proposed ensemble modeling cross-project-based learning approach. Their results demonstrated that their proposed approach surpassed both the Bellwether approach and within-project classification approaches in terms of F-score. 

Afric et al. \cite{B18} focused on classifying bugs into bugs and non-bugs using bug title and description, comparing different approaches such as Simple Keyword Matching (KWM), Improved KWM (IKWM), FastText \cite{FastText}, and RoBERTa \cite{Roberta}. Experimental results revealed that RoBERTa outperformed the other classifiers. Similarly, Terdchanakul et al. \cite{B22} employed the textual contents from bug reports to classify bugs as either bugs or non-bugs. They compared two embedding techniques, namely N-gram IDF and LDA \cite{LDA}, as well as two ML models, LR and RF. Their studies found that the combination of n-gram IDF embeddings with LR achieved the best performance.

\subsection{Bug Prioritization} 
Previous studies also have investigated how to use the textual information of bug reports to predict the priority of bugs by building classification models. Alazzam et al. \cite{B4} utilized bug summary text along with non-textual features such as resolution, product, and the number of comments on bug reports. They proposed a hybrid approach for embeddings using a combination of LDA, Term-to-Term correlation, and neighborhood graph relations, comparing various classifiers. Their findings indicated that the hybrid approach combined with RF and SVM yielded the best results in solving the classification problem of bug priority classes.

Similarly, Izadi et al. \cite{B13} employed bug title and description text to predict bug priority. They used TF-IDF and SentiStrength \footnote{http://sentistrength.wlv.ac.uk/} as their natural language processing techniques and found that RoBERTa outperformed K-Nearest Neighbor (KNN), LR, Multinomial Naive Bayes (MNB), and RF. Ahmed et al. \cite{B28} implemented a tool called Capbug that utilized bug summaries to classify bugs into priority classes. They employed TF-IDF and compared the performances of Naive Bayes (NB), RF, Decision Tree (DT), and LR. Their reported results indicated that RF performed the best among the considered machine learning models when considering the textual features. In a similar vein, Guo et al. \cite{B9} conducted research to classify bugs based on their severity using bug summary as the textual input. They compared models such as NB, MNB, SVM, KNN, DT, and Random Tree. Their findings demonstrated that SVM outperformed the other models.

\subsection{Bug Root Cause} 
A considerable body of literature has also explored automatic bug classification based on the bugs' root causes. Catolino et al. \cite{B15} utilized bug summaries to compare three embedding techniques i.e., TF-IDF, Word2Vec, and Doc2Vec, and experimented with multiple ML models like NB, SVM, LR and RF. The paper \cite{B15} found that the combination of TF-IDF and LR demonstrated the best performance. Ahmed et al. \cite{B28} also used their tool CapBug for bug categorization based on root causes and found that textual features worked best with RF and TF-IDF. \\


The primary focus of our work is centered around the identification of intrinsic bugs based on textual information from bug reports. With the aim of supporting bug prediction models to have a better representation of the real bugs. We argue that incorporating the knowledge of bug report text, as demonstrated in prior studies, can aid in the development of automated approaches for intrinsic bug identification. In our experiments, we employed NLP and ML techniques that have been proven effective in previous studies to construct classification models utilizing the title or the bug description extracted from bug reports.

\section{Approach}
\label{approach}
We now explain the approach we followed. It includes data collection, data pre-processing, embeddings techniques, 
and class imbalance. 


\subsection{Data Collection}
The dataset used in this research was sourced from the reproducibility package included in the study by Rodriguez-Perez et al. \cite{Gema_watchout}. The authors manually classified bug reports into intrinsic bug, extrinsic bug, and no-bug types. The replication package included a wealth of information for each bug report, such as the bug ID, the project name, and the presence of a bug-inducing commit. The dataset contains 1880 bug reports from OpenStack\footnote{https://www.openstack.org}. 

Despite the dataset's comprehensive nature, it lacked crucial textual information about the bug's description, which is necessary for our experimental analysis. Hence, we collected the textual information of bug reports from OpenStack issue tracking system (Launchpad\footnote{https://launchpad.net/openstack}) using a bug-scraping script, which is included in our replication package. Then, we added this information to the dataset. Our extended dataset includes the following information:

\begin{itemize}
    \item Bug ID: The unique identifier for each bug report. 
    \item Project: The name of the project the bug belongs to.
    \item isBUG: Boolean values of 0 or 1 to represent whether the bug report was actually a bug or not based on manual data analysis. 
    \item hasBIC: Boolean values of 0 or 1 to represent whether the bug report can be traced back to a Bug Inducing Commit (BIC) or not. This is the major indicator of whether the bug was an intrinsic bug or an extrinsic bug.
    \item Description: The entire bug description posted in the bug report. 
    \item Title: The bug report's title.
\end{itemize}

\begin{table}
\centering 
\caption{Classification logic}
\label{table: boolean}
\begin{tabular}{l | c| c |c}
\textbf{Case} & \textbf{isBug} & \textbf{hasBIC} & \textbf{Classification}\\ \hline

1 & 1 & 1 & Intrinsic \\ \hline
2 & 1 & 0 & Extrinsic \\ \hline
3 & 0 & X & Non Bug\\ \hline
\end{tabular}
\end{table}

It is worth noting that in order to differentiate the classification of each bug report, boolean conditions utilizing the isBug and hasBIC values were employed to assign labels within the overall dataset. The two boolean values of isBug and hasBIC had been curated as part of the rigorous and manual data analysis performed by Rodriguez-Perez et al. 
 \cite{Gema_watchout}. The classification logic is presented in Table \ref{table: boolean}. When the value of isBug is $1$ and hasBIC is also $1$, it indicates an intrinsic bug, whereas any other combination denotes a non-intrinsic type. Following the application of this labeling logic, out of a total of 1880 bug reports, the dataset contained 1120 intrinsic bugs and 760 non-intrinsic bugs (212 extrinsic bugs and 548 non-bugs). We grouped extrinsic and non-bugs together to form the category ``non-intrinsic'' as our aim is to identify intrinsic bugs. Previous studies have reported that when collecting bug reports from issue tracking systems, some issue reports can be mislabeled~\cite{herzig2013s}, i.e., issue reports that describe defects but were not classified as such (or vice versa).


\subsection{Data Pre-Processing} 
The main data used in our experiments consist of the textual information found in bug titles and descriptions. Upon analysis, we observed 
that the text comprised natural language containing details on bug replication and bug locations, snippets of faulty code, an assortment of technical and project-specific terminology, and lengthy error logs. In order to work with more meaningful text, we applied a comprehensive data cleanup process. This involved the removal of traceback error log data, alphanumeric SHA codes, and commit IDs, as well as special characters and numerical values. URLs and hyperlinks were replaced with the string ``<URL>'' to signify the presence of a URL. All characters were converted to lowercase, and project names were replaced with either ``<internal project>'' or ``<external project>'' depending on whether the bug belonged to the respective project. Stopwords were subsequently eliminated using the Gensim library's standard functionality. Finally, we applied tokenization to transform each segment of text into an iterable list of tokens, and lemmatization to reduce each word to its fundamental root form.

\subsection{Embedding Techniques}
In our study, we generated word embeddings using the seBERT model~\cite{seBERT} and the widely adopted TF-IDF method. SeBERT \cite{seBERT} model was proposed as a software engineering (SE)-oriented variant of the BERT model~\cite{BERT}, specifically pre-trained on a large-scale SE dataset for contextual relevance \cite{seBERT}. Von der Mosel et al. \cite{seBERT} employed 119.7 Gigabytes of processed textual data extracted from diverse sources such as Stackoverflow posts, GitHub issues and commit messages, and JIRA issues to create the SE-context-based dataset used for pretraining BERT. In addition to seBERT, we also employed TF-IDF as an alternative embedding technique for comparison. TF-IDF, which stands for Term Frequency-Inverse Document Frequency, has been extensively utilized in previous bug classification studies~\cite{B4, B13, B14, B15, B17, B21, B28}. It quantifies the importance of a word within a document collection or corpus based on its frequency in the document and its frequency across other documents. 


\subsection{Class Imbalance}
In order to address the issue of class imbalance within the dataset, specifically the disproportionate representation of intrinsic (1120) and non-intrinsic (760) bug classes, we employed the Synthetic Minority Oversampling Technique (SMOTE) \cite{SMOTE}. SMOTE is an effective technique for addressing class imbalance in machine learning tasks and has been used in previous studies on bug classification to handle class imbalance~\cite{B13,B28}. It generates synthetic examples by interpolating between neighboring instances in the feature space. This approach allows for the creation of diverse and representative synthetic examples, effectively expanding the minority class and promoting a more balanced training set. 
Consequently, SMOTE helps in mitigating the risk of the classifier being biased towards the majority class and improves its ability to correctly classify minority class instances. 

\section{Experimental Setup}
\label{experiments}
In this section, we first introduce the ML models studied. Then, we present the key parameter settings of our models. Finally, we describe the evaluation metrics we adopt.

\subsection{Machine Learning Models}
In this study, we employed a selection of five well-established machine learning models. We implemented those models using the open source data analysis library scikit-learn library\footnote{url{https://scikit-learn.org/stable/supervised\_learning.html}} in Python.\\
\textbf{Support Vector Machine (SVM) \cite{SVM}}: SVM is a supervised learning algorithm used for classification and regression tasks. It constructs a hyperplane or a set of hyperplanes in a high-dimensional feature space to separate different classes. SVM aims to maximize the margin between classes, making it effective for handling both linearly separable and non-linearly separable datasets.\\
\textbf{Random Forest \cite{RF}}: It is an ensemble learning method that combines multiple decision trees to make predictions. It creates a collection of decision trees by randomly selecting subsets of features and training each tree independently. During prediction, the final result is determined by aggregating the predictions of individual trees. Random Forest is known for its robustness against overfitting and its ability to handle high-dimensional datasets.\\
\textbf{K-Nearest Neighbors (KNN) \cite{KNN}}: KNN is a non-parametric classification algorithm that assigns a new data point to the majority class of its nearest neighbors. It works by calculating the distances between the new data point and all existing data points in the training set. The value of ``K'' determines the number of neighbors considered. KNN is simple to implement and suitable for small to medium-sized datasets.\\
\textbf{Decision Tree \cite{DT}}: It is a flowchart-like tree structure where each internal node represents a feature or attribute, each branch represents a decision rule, and each leaf node represents the outcome or class label. It partitions the dataset based on the feature values to recursively build a tree-like structure.\\ 
\textbf{Logistic Regression \cite{LR}}: It is a statistical model used for binary classification. It models the relationship between the input features and the probability of belonging to a specific class using the logistic function. LR estimates the coefficients of the input features to make predictions. It is widely used due to its simplicity, interpretability, and ability to provide probabilistic predictions.

These models have been widely utilized within the research community for addressing similar problems~\cite{B2, B5, B13, B4, B13, B28, B9, B10, B12, B17, B22, B15, B20, B14}. The independent variable in our analysis was the textual feature, specifically the bug report title or description, while the dependent variable corresponded to the bug class, either intrinsic or non-intrinsic. To optimize the performance of each model, we utilized the Grid Search optimization algorithm \cite{Grid_Search} for hyperparameter tuning. This approach allowed us to systematically explore various combinations of hyperparameters and select the optimal configuration for each model. The hyperparameter results are described in the next subsection. For robust evaluation and to account for potential biases, we employed the Stratified k-fold cross-validation method with a value of k set to 5. Stratified k-fold cross-validation is a widely used technique for evaluating the performance of machine learning models. In this approach, the original dataset is divided into k equally sized folds while maintaining the proportion of instances from each class in every fold. Specifically, stratification ensures that each fold contains a representative distribution of the different classes present in the dataset. By doing so, it helps to mitigate the potential bias that may arise from imbalanced class distributions. In our study, we employed a value of k=5, meaning that the dataset was divided into five folds, allowing us to perform model training and evaluation five times, with each fold serving as a testing set once while the remaining folds were used for training. 

\subsection{Hyperparameters Selection}
Each ML model was configured with a specific set of hyperparameters, some of which were common across models, while others were model-specific.

For SVM, the primary hyperparameter considered was the kernel type. SVM utilizes mathematical functions for modeling known as kernels, and we experimented with four types: linear, polynomial, radial basis function (RBF), and sigmoid. Additionally, we employed the class weight hyperparameter to assign custom weights to each class in the dataset. Class weights hyperparameter was employed to ensure that the model's training process accounted for the imbalanced distribution of intrinsic and non-intrinsic bugs, enabling it to achieve a more robust and balanced performance across the different bug classes. We explored three possibilities for class weights: balanced, 0.6, and 0.4 for intrinsic and non-intrinsic bug classes and reverse weights of 0.4 and 0.6 for intrinsic and non-intrinsic bug classes. Furthermore, we adjusted the gamma hyperparameter, which defines the kernel coefficient for RBF, polynomial, and sigmoid kernels. Gamma determines the influence of a single training example, with low values indicating a wider influence range and high values indicating a closer influence range. We tested two values for gamma: auto and scale. 

LR was configured with four hyperparameters. The inverse of regularization strength, denoted as C, was set to two values: 1.0 and 0.1. Similar to SVM, LR employed class weights with the same range of possibilities. The penalty hyperparameter imposed a penalty on the logistic model to reduce the influence of less contributive variables, and we experimented with two penalty options: l1 and l2. Two solvers, namely liblinear and lbfgs, were chosen for the optimization problem of LR. The liblinear solver utilizes a coordinate descent algorithm, while the lbfgs solver approximates the Broyden–Fletcher–Goldfarb–Shanno algorithm.

DT and RF shared several hyperparameters. The criterion hyperparameter determined how split impurity would be measured during the tree-building process in DT. We set the maximum depth of the tree, minimum samples required in leaf nodes (min\_samples\_leaf), and minimum samples required to split (min\_samples\_split) an internal node, all within the range of {1, 2, 3} for both DT and RF. 

Lastly, for KNN, we adjusted three hyperparameters: metric, number of neighbors, and weights. The metric hyperparameter allowed us to define the distance metric for computing similarity between neighbors, and we considered ``euclidean'' and ``manhattan'' options. The number of neighbors determined the count of neighbors checked during the classification of a query record, with values ranging from 1 to 3. The weights hyperparameter defined the weight function used, with options including ``uniform'' and ``distance''. The uniform weight function assigned equal weight to all points in the neighborhood, while the distance weight function assigned higher weight to nearby points and lower weight to farther points.

\subsection{Evaluation Metrics}
In order to comprehensively assess the performance of our model, we employed four widely recognized evaluation metrics: precision, recall, accuracy, and F1-score. 

Precision measures the proportion of correctly predicted positive instances (true positives) out of all instances predicted as positive (true positives + false negatives). It quantifies the model's ability to avoid false positives and provides insight into the accuracy of positive predictions.

\begin{equation}
     Precision = \frac{TP}{TP+FP} 
\end{equation}

Recall, also known as sensitivity or true positive rate, calculates the proportion of correctly predicted positive instances (true positives) out of all actual positive instances (true positives + false negatives). It indicates the model's ability to identify all positive instances without missing any.

\begin{equation}
     Recall = \frac{TP}{TP+FN}
\end{equation}

Accuracy calculates the ratio of correctly classified instances (true positives + true negatives) to the total number of instances. It represents the overall correctness of the model's predictions.
\begin{equation}
     Accuracy =  \frac{TP+TN}{TP+TN+FP+FN}
\end{equation}

F1-score is the harmonic mean of precision and recall. It combines both metrics to provide a single value that balances precision and recall. The F1-score is useful when there is an uneven class distribution and serves as a balanced measure of the model's performance.

\begin{equation}
    F1 = \frac{2*Precision*Recall}{Precision+Recall} = \frac{2*TP}{2*TP+FP+FN}
\end{equation}

Each metric is represented as a value between 0 and 1, with higher scores indicating better performance.

In addition, we conducted an evaluation of the Area Under the ROC Curve (AUC-ROC) score for each of our experiments. The AUC-ROC score provides insights into the classification potential of our binary classification model. By examining the AUC-ROC score, we gain a comprehensive understanding of the model's ability to discriminate between intrinsic and non-intrinsic bugs. 

\section{Experimental Results}
\label{results}
In this section, we present the motivation, approach, and results for the following two research questions:


\subsection{RQ1: How does the performance of various NLP and ML techniques differ in accurately identifying intrinsic bugs from bug reports title?}
\textbf{Motivation}: Using just the title of a bug report instead of the body description for training a classification model using NLP techniques might offer several benefits. The title provides a concise summary of the bug's key characteristics, simplifying the input data and reducing computational complexity. Processing only the title is also more efficient, enabling faster model training. Additionally, focusing on the title might help filter out noise from verbose descriptions or irrelevant details present in the bug report descriptions. However, it's essential to consider the trade-offs and limitations of using only the title, as the body description may provide more detailed information and contextual clues that might contribute to accurate bug classification.

\textbf{Approach}: The independent variable for RQ1 is the textual information found in the bug report titles. We applied the pre-processing procedures outlined in the approach section to the independent variable. The dependent variable, indicating the bug class label (intrinsic or non-intrinsic), was transformed into boolean values, with intrinsic bugs assigned a label of 1 and non-intrinsic bugs assigned a label of 0. Following this, we employed our selected embedding techniques (seBERT and TF-IDF) to encode the independent variable. To ensure the availability of training and testing data, we performed an 80-20 split on the embedded data. Subsequently, we applied SMOTE to deal with the class imbalance in the training data. Finally, each ML model was trained using the Grid Search algorithm, which explored all possible combinations of pre-defined hyperparameters for each model. The training process employed a stratified k-fold cross-validation technique with k set to 5. To evaluate the performance of each model, we employed our evaluation metrics: precision, recall, F1-score, accuracy, and AUC-ROC.

\textbf{Results}: The experiments conducted with our implemented automatic bug classification model to categorize bugs into intrinsic and non-intrinsic categories using bug report title are presented in Table \ref{tab: results title}. We identified the optimal hyperparameter combinations for each ML model in our experiments. When utilizing seBERT embeddings, SVM achieved the best performance in terms of F1-score (77\%) with the polynomial kernel, an automatic gamma value, and class weights of 0.6 for intrinsic bugs and 0.4 for non-intrinsic bugs. LR yielded the highest F1-score (73\%) with a C value of 0.1, identical class weights as SVM, an l2 penalty, and the ``liblinear'' solver. DT exhibited its best F1-score (67\%) with the ``gini'' criterion, a maximum tree depth of 2, a minimum number of samples in leaf nodes set to 1, and a minimum number of samples required to split an internal node set to 2. For RF, the best F1-score (71\%) was obtained when the hyperparameters were a maximum tree depth and minimum samples in leaf nodes both set to 3, and a minimum number of samples required to split an internal node set to 2. KNN performed optimally (34\%) with the Manhattan metric, a single neighbor value, and uniform weights.

In contrast, when employing TF-IDF embeddings, SVM demonstrated its highest F1-score (74\%) when using balanced class weights, an automatic gamma value, and the sigmoid kernel. LR achieved its best performance (74\%) with a C value of 1.0, class weights of 0.6 for intrinsic bugs and 0.4 for non-intrinsic bugs, an l1 penalty, and the ``liblinear'' solver. DT achieved its optimal F1-score (78\%) with the ``gini'' criterion, a maximum tree depth of 3, a minimum number of samples in leaf nodes set to 1, and a minimum number of samples required to split an internal node set to 2. RF performed best (77\%) with a maximum tree depth of 3, a minimum number of samples in leaf nodes set to 1, and a minimum number of samples required to split an internal node set to 3. Similarly, KNN exhibited its best F1-score (64\%) with the Euclidean metric, a single neighbor, and uniform weights. 

In terms of the AUC-ROC, the experimental results utilizing seBERT embeddings reveal that SVM and RF achieve the highest scores with 66\%. They are closely followed by DT with a score of 63\%, LR with a score of 62\%, and finally, KNN with a score of 56\%. When employing TF-IDF embeddings, the experimental outcomes demonstrate that LR achieves the highest AUC-ROC score of 64\%. It is followed closely by RF with a score of 63.8\%, followed by SVM with a score of 63\%. DT obtains a score of 61\%, while KNN attains the lowest score of 58\%.

\begin{tcolorbox}[colback=gray!8!white,colframe=gray!75!black]
The results show that for bug report titles collected from OpenStack when utilizing the seBERT embedding technique, SVM achieved the highest F1 score (77\%) as well as AUC-ROC score (66\%), sharing the same highest AUC-ROC score with RF. 
Among the experiments conducted using TF-IDF, DT achieved the highest F1 score (78\%), slightly outperforming the highest of seBERT combined with SVM, which achieved 77\%.
\end{tcolorbox}

\begin{table*}
  \caption{Results of experiments conducted with Title of Bug Reports}
  \label{tab: results title}
  \begin{tabular}{{p{0.2\textwidth}|
  p{0.08\textwidth}|
p{0.06\textwidth}| p{0.07\textwidth}|p{0.08\textwidth}|
p{0.08\textwidth}| p{0.06\textwidth}| p{0.07\textwidth}|p{0.08\textwidth}}}
    \multicolumn{1}{c|}{} & \multicolumn{4}{c|}{seBERT}
     & \multicolumn{4}{c}{TF-IDF}\\ \hline
     \toprule
    Models & Precision & Recall & F1 Score &AUC-ROC&Precision & Recall & F1 Score&AUC-ROC\\ \hline
    Support Vector Machine & 71\% & \textbf{83\%} & \textbf{77\%} &  \textbf{66\%}
    & 69\% & 80\% & 74\% & 63\% \\\hline
    Logistic Regression & 68\% & 79\% &  73\%  & 62\%
    & \textbf{70\%} & 79\%  &  74\% &  \textbf{64\%}    \\\hline
    Decision Tree & 71\%  &  63\% &  67\%   & 63\%
    & 65\% & \textbf{96\%} &  \textbf{78\%}  & 61\%\\\hline
    KNearest Neighbor &  \textbf{78\%} & 22\%  & 34\% &  56\%      
    & 67\% & 61\% & 64\% & 58\%\\\hline
    Random Forest & 73\%  &  70\% &  71\%   & \textbf{66\%}
    & 68\% & 89\% & 77\% & 63.8\%\\\hline
    \bottomrule
\end{tabular}
\end{table*}

\subsection{RQ2: How does the performance of various NLP and ML techniques differ in accurately identifying intrinsic bugs from bug report description?}
\textbf{Motivation}: In light of the potential benefits associated with the detailed information and contextual clues present in bug descriptions, our study aims to investigate the performance of the approaches utilized in RQ1 when utilizing longer textual information from bug descriptions. Accurate identification of intrinsic bugs is crucial for software quality, as it helps to build reliable datasets to improve bug prediction models' performance and increase their trustworthiness. 

\textbf{Approach}: To answer RQ2, we use the description of bug reports as our independent variable. Then, we followed the same approach described in RQ1 and repeated all the experiments.

\textbf{Results}: The experiments conducted with our implemented automatic bug classification model to categorize bugs into intrinsic and non-intrinsic categories using bug report description are presented in Table \ref{tab: results desc}. We identified the optimal hyperparameter combinations for each ML model in the experiments utilizing seBERT embeddings as well. LR yielded the best F1-score (76\%) with a C value of 1.0, class weights of 0.6 for intrinsic bugs and 0.4 for non-intrinsic bugs, an l1 penalty, and the ``liblinear'' solver. SVM achieved its highest performance in terms of F1-score (75\%) with the polynomial kernel, gamma value of ``scale'', and class weights identical to LR. For RF, the best F1-score (68\%) was obtained when the hyperparameters were a maximum tree depth of 3 and minimum samples in leaf nodes set to 1, and a minimum number of samples required to split an internal node set to 2. DT exhibited its best performance (49\%) with the ``gini'' criterion, a maximum tree depth of 3, a minimum number of samples in leaf nodes set to 1, and a minimum number of samples required to split an internal node set to 2. KNN performed optimally (11\%) with the Manhattan metric, a single neighbor value, and uniform weights.

In contrast, when employing TF-IDF embeddings, LR again yielded the best F1-score (75\%) with a C value of 1.0, balanced class weights, an l1 penalty, and the ``liblinear'' solver. Following LR closely, SVM achieved its highest performance in terms of F1-score (74\%) with the sigmoid kernel, gamma value of ``scale'', and class weights identical to LR. For RF, the best F1-score (67\%) was obtained when the hyperparameters were a maximum tree depth of 3 and minimum samples in leaf nodes set to 1, and a minimum number of samples required to split an internal node set to 3. DT exhibited its best performance (37\%) with the ``entropy'' criterion, a maximum tree depth of 3, a minimum number of samples in leaf nodes set to 3, and a minimum number of samples required to split an internal node set to 2. KNN performed optimally (35\%) with the Euclidean metric, a single neighbor value, and uniform weights. 

In terms of the AUC-ROC, the experimental results utilizing seBERT embeddings reveal that LR achieves the highest scores of 65.2\%, which is closely followed by SVM with a score of 65\%. RF exhibits an AUC-ROC score of 61.8\%, DT scores 54.4\%, and finally, KNN yielded a score of 51.2\%. When employing TF-IDF embeddings, the experimental outcomes demonstrate the same order from best to worst AUC-ROC scores. LR achieves the highest AUC-ROC score of 66\%, followed by SVM with a score of 65.4\%, and then by RF with a score of 65.1\%. DT obtains a score of 55.4\%, while KNN attains the lowest AUC-ROC score of 52.6\%.

\begin{tcolorbox}[colback=gray!8!white,colframe=gray!75!black]
The results indicate that for bug report descriptions collected from OpenStack when using seBERT, LR outperforms all other models in terms of F1-sore (76\%) as well as in terms of AUC-ROC score (65.2\%). 
Also, when using TF-IDF, the highest scoring models remain the same, LR demonstrating the best F1-score (75\%) and AUC-ROC score (66\%). 
\end{tcolorbox}


\begin{table*}
  \caption{Results of experiments conducted with Description of Bug Reports}
  \label{tab: results desc}
  \begin{tabular}{{p{0.2\textwidth}|
  p{0.08\textwidth}|
p{0.06\textwidth}| p{0.07\textwidth}|p{0.08\textwidth}|
p{0.08\textwidth}| p{0.06\textwidth}| p{0.07\textwidth}|p{0.08\textwidth}}}
    \multicolumn{1}{c|}{} & \multicolumn{4}{c|}{seBERT}
     & \multicolumn{4}{c}{TF-IDF}\\ \hline
     \toprule
    Models & Precision & Recall & F1 Score &AUC-ROC&Precision & Recall & F1 Score&AUC-ROC\\ \hline
    Support Vector Machine & 70\% & 81\% & 75\% & 65\%
    & 71\% & 77\% & 74\% & 65.4\% \\\hline
    Logistic Regression & 70\% & \textbf{83\%} &  \textbf{76\% } & \textbf{65.2\%}
    & 71\% & \textbf{78\%}  &  \textbf{75\%} &  \textbf{66\%}    \\\hline
    Decision Tree & 66\%  &  39\% &  49\%   & 54.5\%
    & 72\% & 25\% &  37\% & 55.4\%\\\hline
    KNearest Neighbor &  \textbf{72\%} & 6\%  & 11\% &  51.2\%      
    & 65\% & 24\% & 35\% & 52.6\%\\\hline
    Random Forest & 70\%  &  67\% &  68\%   & 61.8\%
    & \textbf{74\%} & 61\% & 67\% & 65.1\%\\\hline
    \bottomrule
\end{tabular}
\end{table*}

\section{Discussion}
\label{discussion}
In this section, we discuss our findings, the limitations we found during our experimental setup, the opportunities to enhance our results in the future, and the implications of our results.

\subsection{Classifying Intrinsic Bugs from Bug Reports' Textual Information}
At the beginning of our study, we posited that leveraging the knowledge of the textual information of bug reports could help develop approaches to automatically identify intrinsic bugs. Our results clearly show that bug reports can be classified into intrinsic and non-intrinsic bugs with a pretty high F1-score (76\% and 77\%) using machine learning models.

Catolino et al. \cite{B15} utilized bug summaries, which were summarized form of bug reports, to classify bugs based on root causes. They compared three embedding techniques: TF-IDF, Word2Vec, and Doc2Vec, and experimented with multiple ML models like Naive Bayes, SVM, LR and RF. They reported that the best performance (64\%, in terms of F1-score) was obtained by using TF-IDF for embedding and LR as the classifier. Our study also indicates that combining TF-IDF and LR when using bug descriptions is the best approach for classifying intrinsic and non-intrinsic bugs, with an F1-score of 75\%. 

The implementation of DeepLabel by Li et al. \cite{B10} resulted in F1-scores of 79\% and 89\% for bug and non-bug classes. Their implementation utilized bug title and description as input, and combined Word2Vec, Attention-based Bi-directional LSTM (ABLSTM) model, and ensemble method to combine title classifier and description classifier together, yielding high results. In our study, we deliberately chose not to combine bug titles and descriptions to explore whether utilizing shorter textual information (bug titles) would yield comparable performance to the classification of intrinsic and non-intrinsic bugs. However, the combination of bug title and bug report is in our future research agenda.

Our findings indicate that using just the title of bug reports might be enough to identify intrinsic bugs, as indicated by the F1-score (78\%) and AUC (65\%). This is a very interesting finding as it implies that using longer text information does not outperform the classification of intrinsic bugs. We hypothesize that our findings might be due to the conciseness present in the title of bug reports. Titles often contain a succinct summary of the issue, providing a condensed representation of the bug's key characteristics. Furthermore, bug report bodies usually can contain verbose descriptions, discussions, or irrelevant details that may introduce noise into the classification model. Focusing solely on the title can help filter out some of this noise. Thus, using only the title can simplify the input data and reduce the computational complexity of the classification model without losing performance. We suggest that if software practitioners have computational problems, they can use only the title of bug reports to classify bugs into intrinsic or non-intrinsic.

\subsection{Limitations and Opportunities}
\textbf{Bug report features}: Although our results indicate a favorable performance of the classification model using the textual information of bug reports (description or title), we have observed instances where its effectiveness is limited due to the insufficiency of words within bug reports for identifying intrinsic and non-intrinsic bug types accurately. Therefore, conducting further studies on the linguistic patterns employed by developers would be valuable in enhancing the classification of bug reports. We also hypothesize that some of the language used in bug reports might be context-dependent, which makes it more difficult for the approach to understand the context.  
Additionally, exploring different combinations of textual features through ensemble learning would be an opportunity for the future. While we have examined bug report titles and descriptions individually, there is potential for enhanced performance by considering them in combination. Analyzing the value of other bug report features, such as developers' discussions, or experimenting with other textual features, such as bug fixing commit messages, and exploring different combinations of these features could provide valuable insights to understand how different classification models perform when classifying intrinsic bugs.

\textbf{Prepossessing techniques}:
In our study, following previous studies, we applied some pre-processing techniques. 
However, there is a need for future investigations to conduct a more comprehensive analysis to understand the specific impact of these pre-processing techniques on the performance of our classification model. To further advance the pre-processing stage, targeted techniques such as code abstraction and selective stopword removal can be considered. These techniques have the potential to refine the textual data by reducing noise and increasing the relevance of extracted features \cite{B5, B10, B21}, ultimately aiming to improve the overall results of the classification model.

\subsection{Implications}
\textbf{Bug prediction models}: 
In this paper, we have demonstrated that it is possible to identify intrinsic bugs using the textual information of bug reports. Our findings show that seBERT and TF-IDF can be effectively utilized for intrinsic bug identification. The highest performance scores (F1 score of 78\%) were achieved by combining TF-IDF with the DT and utilizing the bug titles. This was closely followed by seBERT, SVM, and bug titles (with an F1 score of 77\%).

By classifying intrinsic bugs and non-intrinsic bugs, our proposed approach aids not only JIT bug prediction models but all bug prediction models that use information from bug-inducing commits to train their models. Such models could achieve higher accuracy when trained solely on intrinsic bugs, providing a more faithful representation of real-world bugs. By incorporating our proposed approach as part of bug prediction practices, bug prediction models can improve their precision in detecting and forecasting intrinsic bugs, thereby enhancing their overall performance. 
Our approach does not mitigate the limitations of relying solely on the SZZ algorithm to link bugs with bug-inducing commits (i.e., false positives~\cite{Gema_watchout,fan2019impact}, refactorings~\cite{neto2018impact}, and false negatives~\cite{rezk2021ghost,sahal2018identifying}). However, previous studies have proposed solutions to control for this noise~\cite{neto2018impact, da2016framework}. Hence, we believe that by combining our approach with these previously proposed approaches, researchers and practitioners can obtain more reliable datasets that can be used to feed bug prediction models. 

\textbf{Devising new tools}: The process of curating bug datasets is resource-intensive, demanding significant effort and expertise in the specific software system. This labor-intensive task might incur substantial costs due to the expertise required, manual efforts involved, and data pre-processing, making it a financially demanding process. Thus the development of tools that help in the classification of bugs might be useful for researchers. In this paper, we proposed an approach that can be integrated into the software engineering process as part of a software tool. This tool could automatically identify intrinsic bugs based on bug report titles or descriptions and flag them so researchers could automatically collect flagged intrinsic bugs when creating bug datasets.

\section{Threats to Validity}
\label{threats}
The validity of this study is described in terms of the three main threats to validity that affect empirical software engineering research: construct, internal, external, and conclusion ~\cite{wohlin2012experimentation}.

\textbf{Construct Validity}. Since we are using the replication package provided in Rodriguez-Perez et al.’s paper~\cite{Gema_watchout}, this study suffers from the same construct validity threats reported in Rodriguez-Perez et al.’s study. From the dataset, 705 issues were analyzed by only a single rater. This might threaten the validity of the results. However, to minimize the impact of this threat, those raters were trained until they achieved a near-perfect agreement before they started classifying the 705 issues. 

\textbf{Internal Validity}. The design of a classification model encompasses numerous subjective decisions, such as balancing the data, tuning the parameters, and validation techniques, among others. These decisions can potentially impact the classification results. To mitigate some of these threats, we follow the suggestions of previous studies and tune the hyperparameters~\cite{fu2016tuning,tantithamthavorn2018impact}. For validating the model, we relied on 5-fold cross-validation as our dataset is small, ensuring sufficient representation of different bug classes while maintaining computational efficiency. Finally, the domain-specifity of OpenStack might put some threats when using seBERT as seBERT has been pre-trained on the systems Github, Jira, and Stack Overflow. However, OpenStack uses Launchpad as its issue-tracking system, so there might be differences in the pattern of those systems that can affect the performance of seBERT in our study.

\textbf{External Validity}. The study of just one project, OpenStack, prevents us from generalizing our findings to other systems. However, our goal was not to claim that our results would stand to all systems but rather to show that by using NLP and ML techniques, we can automatically identify intrinsic bugs. We think that our research is successful in this regard, as we demonstrate that intrinsic bugs can be identified using different techniques. We offer sufficient evidence that researchers and practitioners can use our proposed approach to identify intrinsic bugs to curate a dataset. Moreover, since OpenStack is considered a mature open-source project, results might not generalize to immature projects. Additionally,
it is possible that other projects may require specific pre-processing steps due to the fact that some of these steps might depend on the domain-specific language of the project.

\textbf{Conclusion Validity}. We evaluate our classification model using a number of different evaluation metrics, such as precision, recall, F1, and AUC-ROC. It is known that AUC-ROC is sensitive to imbalanced data. Since our dataset is unbalanced, we used SMOTE to balance the data.


\section{Conclusion and Future work}
\label{conclusion}
Currently, there is a lack of automated approaches to assist practitioners in the challenging task of classifying intrinsic and extrinsic bugs. Addressing this gap, our study aims to bridge this gap by introducing the first approach that classifies bug reports as either intrinsic or non-intrinsic (i.e., extrinsic and non-bugs) using their textual information. Furthermore, we conducted a comprehensive set of experiments to investigate the effectiveness of various combined techniques and explore the impact of the length of textual information on classification performance.

Our results indicate that 
both seBERT and TF-IDF are effective in identifying intrinsic bugs. The best performance was observed when combining TF-IDF with the Decision Tree algorithm and utilizing bug titles, resulting in an  F1 score of 78\%. Following closely, seBERT, Support Vector Machine, and bug titles achieved a strong F1 score of 77\%. Furthermore, our findings suggest that incorporating longer textual information, specifically the bug reports' descriptions, does not yield superior performance compared to models utilizing shorter information, such as bug report titles.

Our future work focuses on improving the devised model and creating a taxonomy for intrinsic and non-intrinsic bugs. There are several potential avenues for further enhancing our classification approach as we have discussed in the limitations and opportunities subsection. First, the utilization of contemporary classifiers, such as Deep Learning models or fine-tuned Large Language models could be explored to increase the performance of our classification model. These advanced techniques might 
yield improved performance when applied to our dataset. Second, it would be beneficial to conduct experiments and compare the effectiveness of non-textual features, such as source code metrics, in classifying bugs into intrinsic and non-intrinsic categories. By examining the efficiency of these features in comparison to textual features extracted from bug reports, we can determine if they offer substantial improvements in bug classification performance. Also, these experiments would help in creating the taxonomy of intrinsic and non-intrinsic bugs. Finally, we will explore the potential benefits of incorporating information regarding the origin of bugs, e.g., external dependencies associated with extrinsic bugs might contribute to their perceived importance in comparison to intrinsic bugs.

\textbf{Replication Package}:
In order to promote reproducible research, the dataset and scripts for this paper are available in our replication package.\footnote{\url{https://zenodo.org/record/8125762}}

\section*{Acknowledgments}
The authors acknowledge the support of the Natural Sciences and Engineering Research Council of Canada (NSERC), funding reference number GR023171, in this project. 


\bibliographystyle{ACM-Reference-Format}
\bibliography{sample-base}

\end{document}